\documentclass[a4paper,11pt]{article}
\usepackage{pos}
\usepackage{url}
\usepackage{subfigure}

\title{The Global Cosmic Ray Observatory -- Challenging next-generation multi-messenger astronomy with interdisciplinary research}
 \ShortTitle{The Global Cosmic Ray Observatory -- Challenging next-generation multi-messenger astronomy}

\author{Toshihiro Fujii on behalf of the GCOS supporters}

\affiliation[]{Graduate School of Science, Osaka Metropolitan University, Sumiyoshi, Osaka 558-8585, Japan}

\affiliation[]{Nambu Yoichiro Institute of Theoretical and Experimental Physics,\\
Osaka Metropolitan University, Sumiyoshi, Osaka 558-8585, Japan}

\emailAdd{toshi@omu.ac.jp}

\abstract{
The origin of ultra-high-energy cosmic rays (UHECRs) is one of the most intriguing mysteries in astroparticle physics and high-energy physics. Since UHECRs with light mass compositions are less deflected by the Galactic and extragalactic magnetic fields, their arrival directions are more strongly correlated with their origins. Charged-particle astronomy with UHECRs is hence a potentially viable probe of extremely energetic phenomena in the universe. The Global Cosmic Ray Observatory (GCOS) is a proposed next-generation observatory to elucidate these origins through precise measurements of UHECRs with unprecedented exposure and mass identification capabilities. We will focus on the ideas and requirements for GCOS summarized in \url{https://arxiv.org/abs/2502.05657} 
and share the recent advances in detector developments and future perspectives with interdisciplinary research.
\\\\\\ 
To be published in PoS (ICRC2025) 258\,\,~\url{https://pos.sissa.it/501/258/}
}

\ConferenceLogo{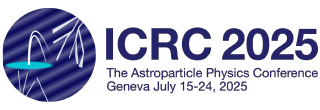}

\FullConference{39th International Cosmic Ray Conference (ICRC2025)\\
 15–24 July 2025\\
Geneva, Switzerland\\}

\begin{document}
\maketitle

\section{Detection of ultra-high-energy cosmic rays}

Motivated by a discovery of an extremely energetic cosmic ray with an energy of 100\,EeV ($\equiv$ 10$^{20}$\,eV) in 1962~\cite{Linsley:1963km}, scientists have been enthusiastically searching for the universe's most energetic particles over 60 years.
The origin and nature of cosmic rays are one of the most important mysteries in astroparticle physics and high-energy physics. 
Ultra-high energy cosmic rays (UHECRs) with energies above $\sim$50\,EeV interact with 2.7\,K cosmic microwave background radiations and extragalactic background light via photo-meson production for proton, and photo-disintegration process for heavy nuclei.
Due to their limited mean free paths, the sources of UHECRs are limited in a cosmological neighbourhood at a distance of 50--100\,Mpc, resulting in a cutoff or suppression of the energy spectrum above $\sim$50\,EeV, dubbed ``GZK cutoff''~\cite{bib:gzk1,bib:gzk2}.
Since UHECRs with light mass composition are less deflected by the Galactic and extragalactic magnetic fields due to their enormous kinetic energies, their arrival directions are more likely to be correlated with their sources than their lower-energy counterparts.
As next-generation astronomy, charged-particle astronomy with UHECRs is a potentially viable probe of extremely energetic phenomena in the nearby universe, as illustrated in Figure~\ref{fig:uhecr_astronomy}A.

In current observatories, two types of detectors are mainly used for UHECR measurements: arrays of ground detectors (e.g. plastic scintillators, water-Cherenkov detectors, and radio detectors) that sample secondary particles from extensive air shower at the ground, and large-field-of-view telescopes that directly measure ultra-violet nitrogen fluorescence photons for calorimetric energy determination. 
The two largest UHECR observatories in operation at both hemispheres, the Pierre Auger Observatory (Auger) in Mendoza, Argentina~\cite{bib:auger}, and the Telescope Array experiment (TA) in Utah, USA~\cite{bib:tafd, bib:tasd}, are hybrid detectors that combine both techniques, employing arrays of ground detectors overlooked by fluorescence telescopes. 

\section{Challenging the next-generation astronomy using UHECRs}
\begin{figure}[t]
    \centering
    \subfigure[Charged-particle astronomy of UHECRs]{\includegraphics[width=0.45\linewidth]{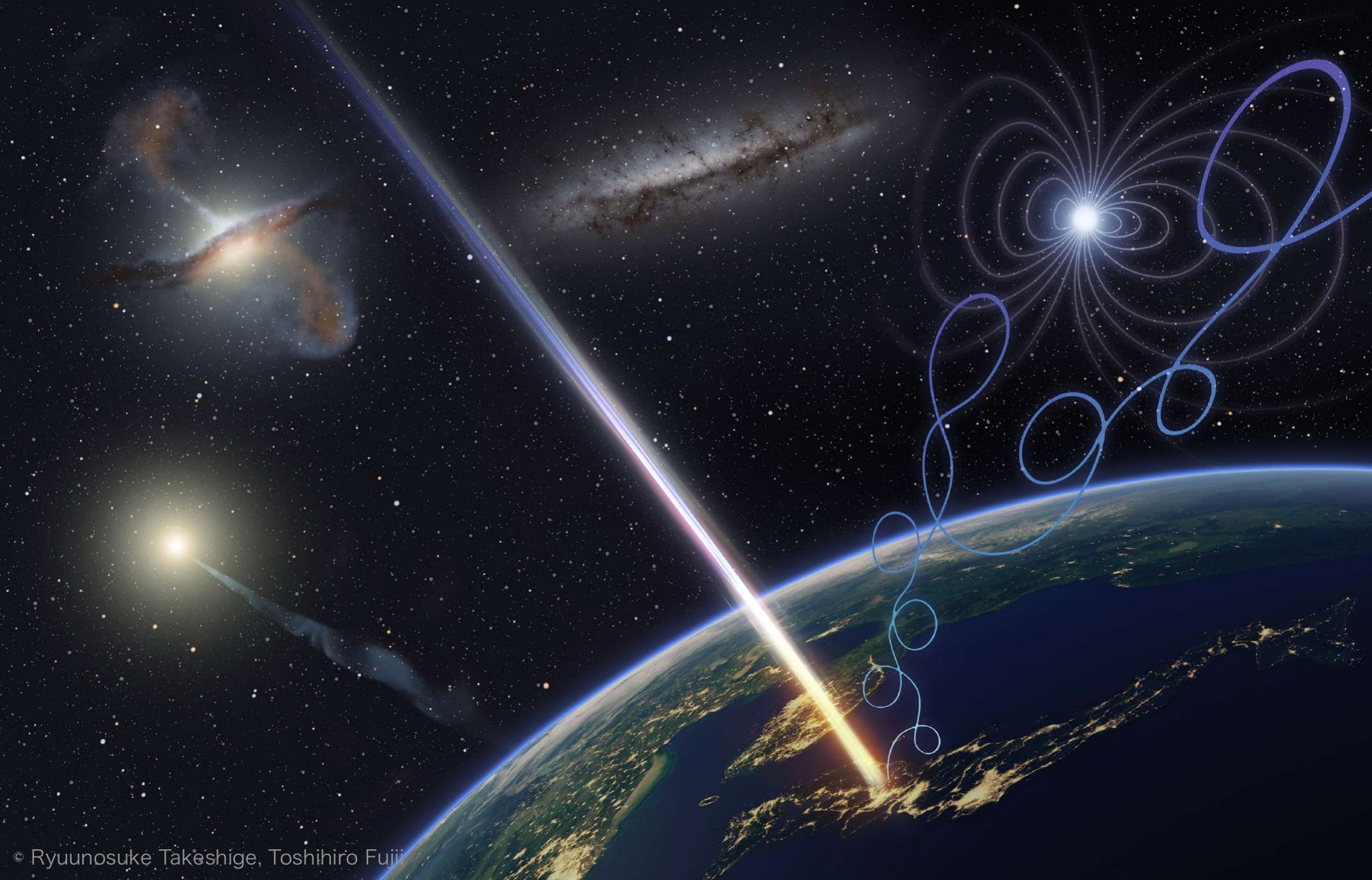}}
    \subfigure[Arrival directions of UHECRs above 100\,EeV]{\includegraphics[width=0.5\linewidth]{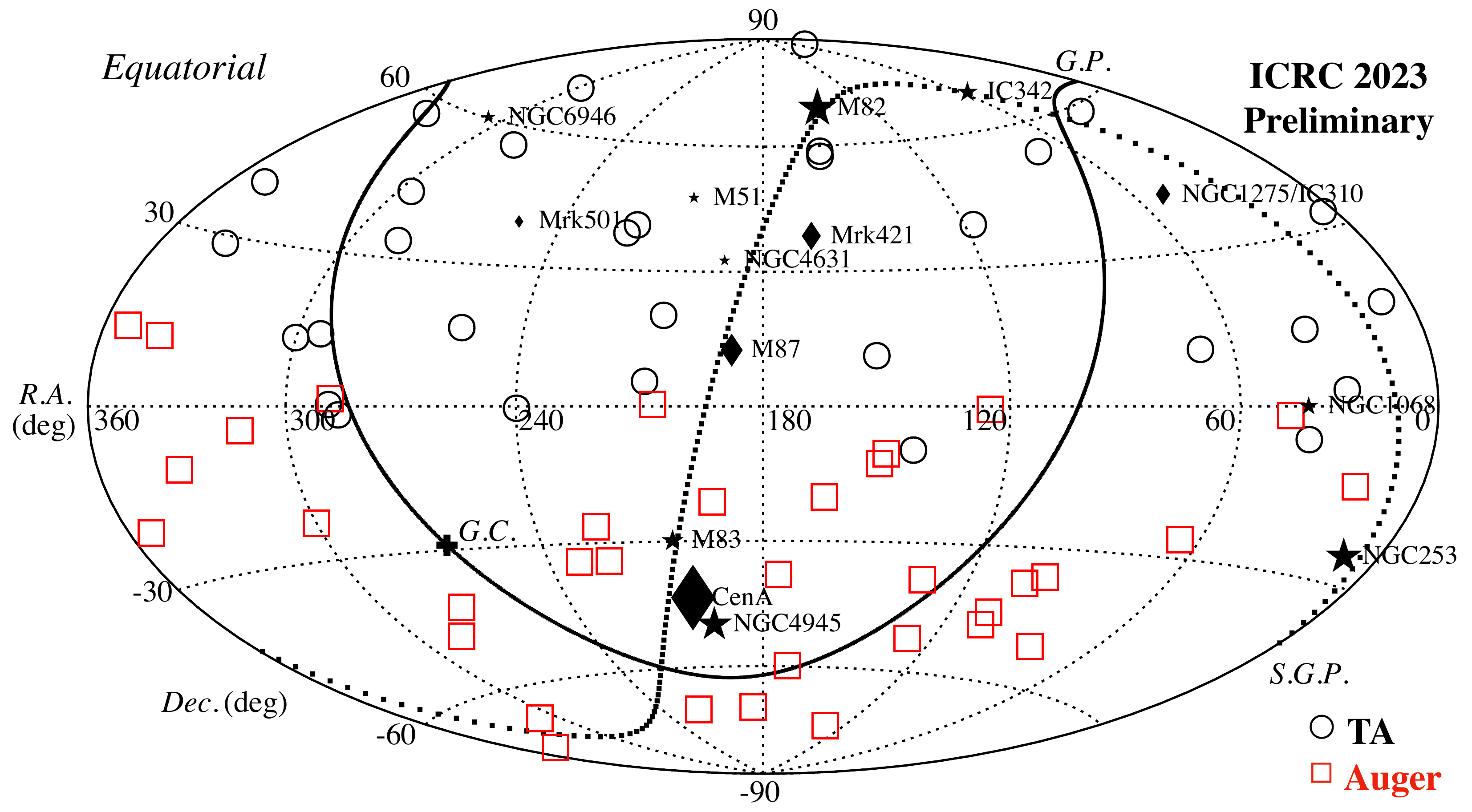}}
    \caption{(A) Conceptual image to indicate UHECR astronomy. The background image shows possible UHECR source candidates, such as active galactic nuclei, starburst galaxies and magnetars. (B) Arrival directions of UHECRs with energies above 100\,EeV measured with Auger (red open square) and TA (black open circle), overlaying the promising nearby source candidates, like active galactic nuclei (filled diamond),  starburst galaxies (filled star). Taken from ~\cite{Fujii:2024sys}. }
    \label{fig:uhecr_astronomy}
\end{figure}

Recent progress and latest results from Auger and TA are summarized in~\cite{Coleman:2022abf}.
The cutoff of the energy spectrum above $\sim$50\,EeV~\cite{PierreAuger:2020kuy,Kim:2025yzr},
a heavy mass composition at the highest energies~\cite{PierreAuger:2024flk,TelescopeArray:2024oux}, and a significant large-scale anisotropy in arrival directions above 8\,EeV~\cite{PierreAuger:2024fgl,PierreAuger:2017pzq} 
were reported.
Figure~\ref{fig:uhecr_astronomy}B shows the arrival directions of UHECRs above 100\,EeV measured with Auger and TA, indicating a nearly isotropic distributions~\cite{Fujii:2024sys}.
No significant clusterings and correlations in a direction of nearby promising source candidates are found in the current statistics above 100\,EeV.
This isotropic distribution may imply heavier mass compositions, stronger Galactic and/or extragalactic magnetic fields, and more abundant source densities than previously expected.

\begin{figure}[b]
    \centering
    \subfigure[Extensive air shower]{\includegraphics[width=0.44\textwidth]{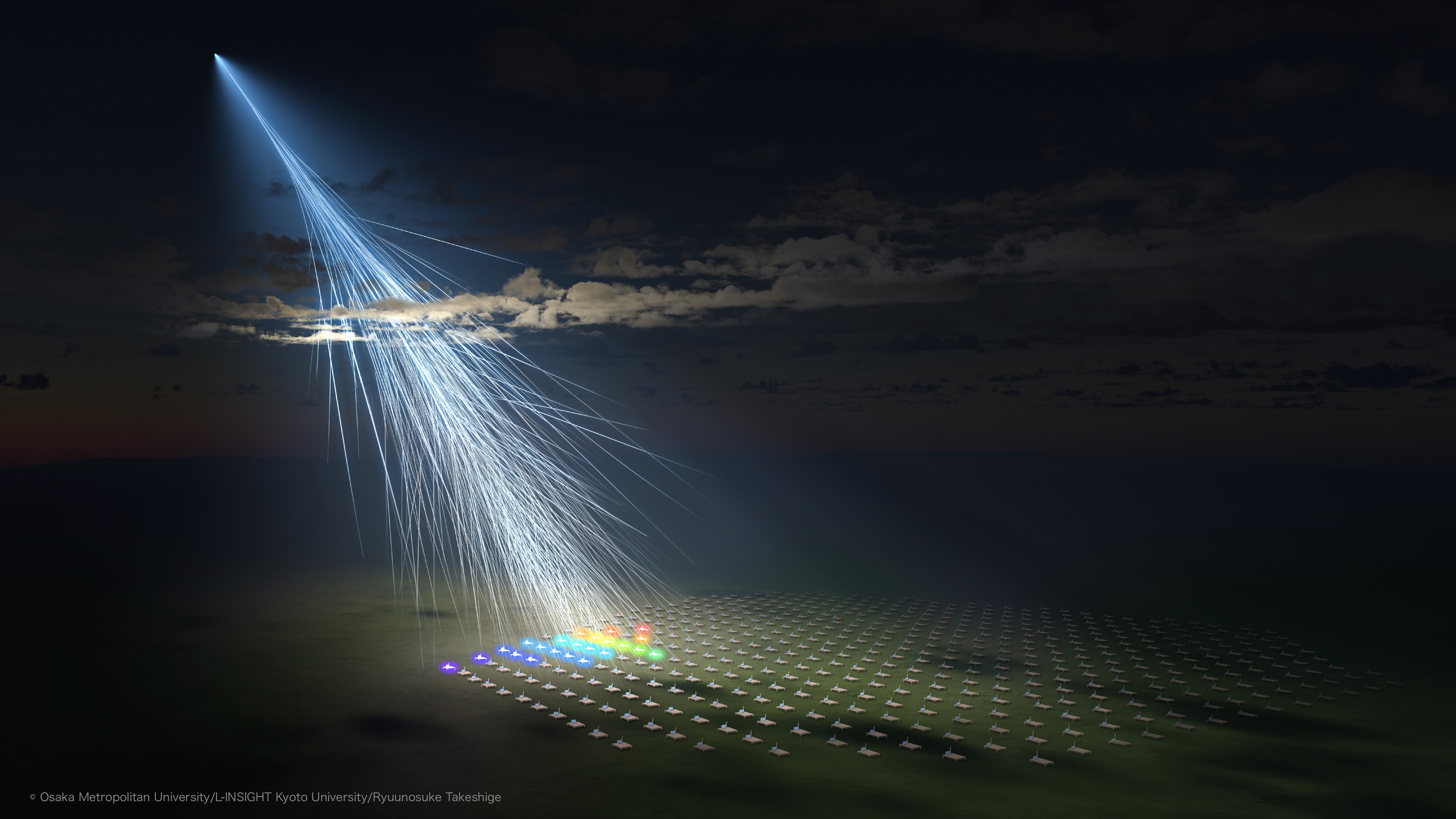}}
    \subfigure[Arrival direction]{\includegraphics[width=0.55\textwidth]{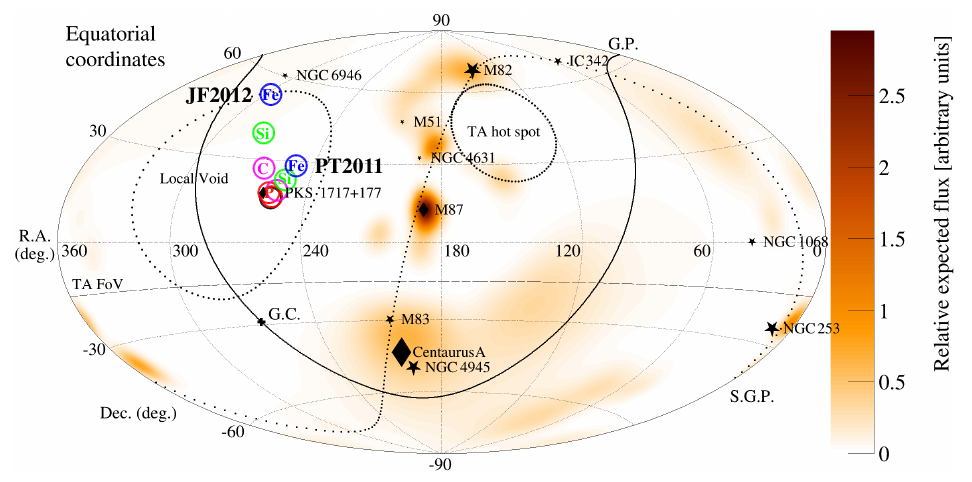}}
    \caption{(A) Conceptual image of the extensive air shower induced by the 244-EeV event detected with TA, dubbed the ``Amaterasu particle'', and (B) its arrival direction in the equatorial coordinates and back-tracking results in the Galactic magnetic field models assuming proton, carbon, silicon and iron primaries. Taken from ~\cite{TelescopeArray:2023sbd}.}
    \label{fig:amaterasu}
\end{figure}

The most energetic cosmic ray measured with the current observatories is 244\,$\pm$\,29\,(stat.) $^{+51}_{-76}$\,(syst.)\,EeV ($\sim$ 40\,joules) detected with TA on May 27th 2021, dubbed the ``Amaterasu particle''~\cite{TelescopeArray:2023sbd}.
Figure~\ref{fig:amaterasu}A shows a conceptual image of the extensive air shower induced by the Amaterasu particle detected with the TA. 
As shown in Figure~\ref{fig:amaterasu}B, the Amaterasu particle came from a direction of the ``Local Void'' even if taking account of a deflection for heavy mass composition in the Galactic magnetic field, far away from the nearby possible source candidates.
As indicated from a large differences of back-tracking results among primary species in the magnetic field models in Figure~\ref{fig:amaterasu}B, the mass identification capability is essential to disentangle origins of UHECRs.
The source region and distance of the Amaterasu particle are addressed to consider the primary species, the Galactic and extragalactic magnetic fields, and energy uncertainties~\cite{Unger:2023hnu,Kuznetsov:2023jfw,Bourriche:2025qnj}.
Possible explanation include a large deflection of ultra-heavy compositions~\cite{Zhang:2024sjp}, origin of binary neutron star merger~\cite{Farrar:2024zsm}, transient phenomena in magnetars~\cite{Shimoda:2024qzw}, and also new physics beyond standard model such as magnetic monopole~\cite{Frampton:2024shp,Cho:2023krz}, superheavy dark matter~\cite{Sarmah:2024ffy,Murase:2025uwv}, 
and quantum gravity~\cite{,Addazi:2021xuf,Lang:2024jmc,Das:2025tfq}.

The current observations are statistically limited due to the cutoff of the energy spectrum.
Next-generation observatory will require an unprecedented exposure, exceeding current observatories by an order of magnitude, and mass identification capabilities.
A worldwide cooperation is essential to construct such a huge detector.

\section{The Global Cosmic Ray Observatory} 

\begin{figure}
    \centering
    \includegraphics[width=0.9\textwidth]{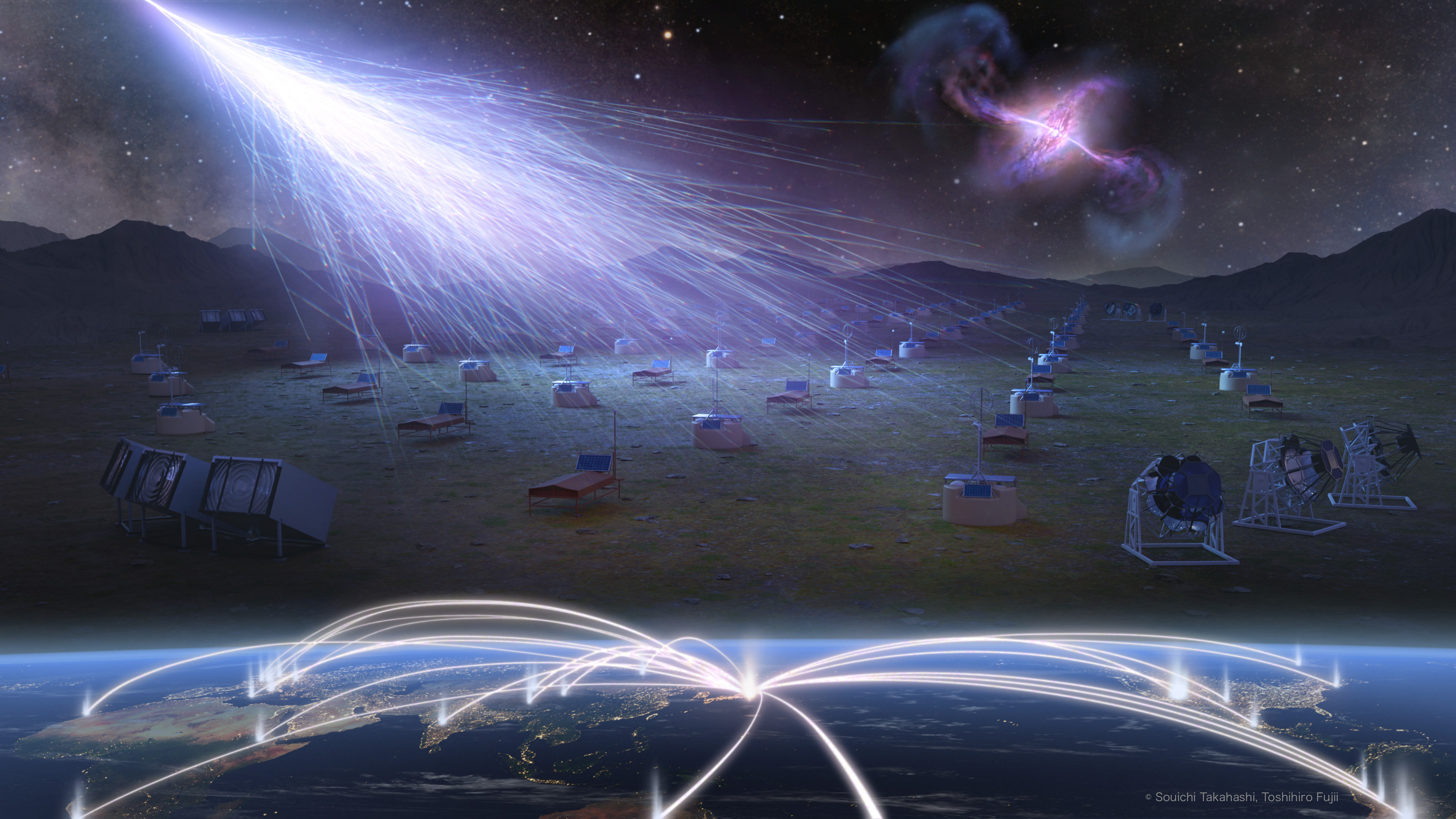}
    \caption{Conceptual image of GCOS for detecting UHECRs through the worldwide cooperation. Taken from the 4th GCOS workshop  \url{https://indico.cern.ch/e/gcos2025}.}
    \label{fig:gcos}
\end{figure}

The Global Cosmic Ray Observatory (GCOS) is a proposed next-generation observatory to elucidate the origin and nature of UHECRs.
Figure~\ref{fig:gcos} shows a conceptual image of GCOS. 
The number of sites for GCOS will be more than two through the worldwide cooperation to achieve a total effective coverage of 60,000\,km$^2$ 
with a trigger threshold of 10\,EeV and a high-quality event threshold of 30\,EeV.
The expected reconstruction accuracies of UHECRs are 1.0$^\circ$ for arrival direction, 10\% for energy, and $\Delta\ln(A)=1.0$ for the logarithmic mass number of primary species.
Several detector designs are under consideration, such as a layered water Cherenkov detector~\cite{Flaggs:2025ofl}, low-cost fluorescence telescope~\cite{YTameda:2025yll,FAST:2025exr}, radio detector~\cite{deErrico:2025cgm}, or their combinations. 
The detector of GCOS will be required to be a high duty-cycle, low maintenance and autonomous operation. 
Ideas and requirements of GCOS are summarized in~\cite{Ahlers:2025pqg}.

\begin{figure}
    \centering
    \subfigure[Total exposure]{\includegraphics[width=0.51\textwidth]{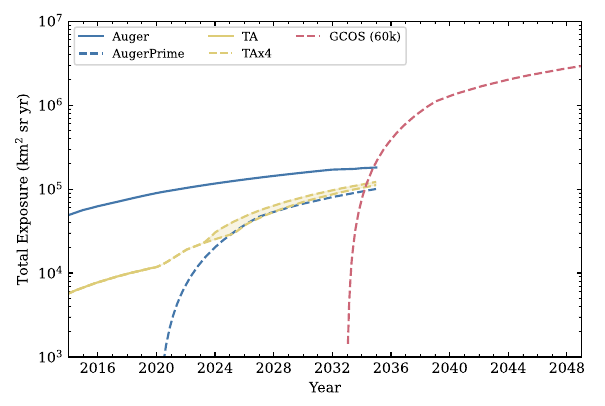}}
    \subfigure[Expected sensitivity]{\includegraphics[width=0.48\textwidth]{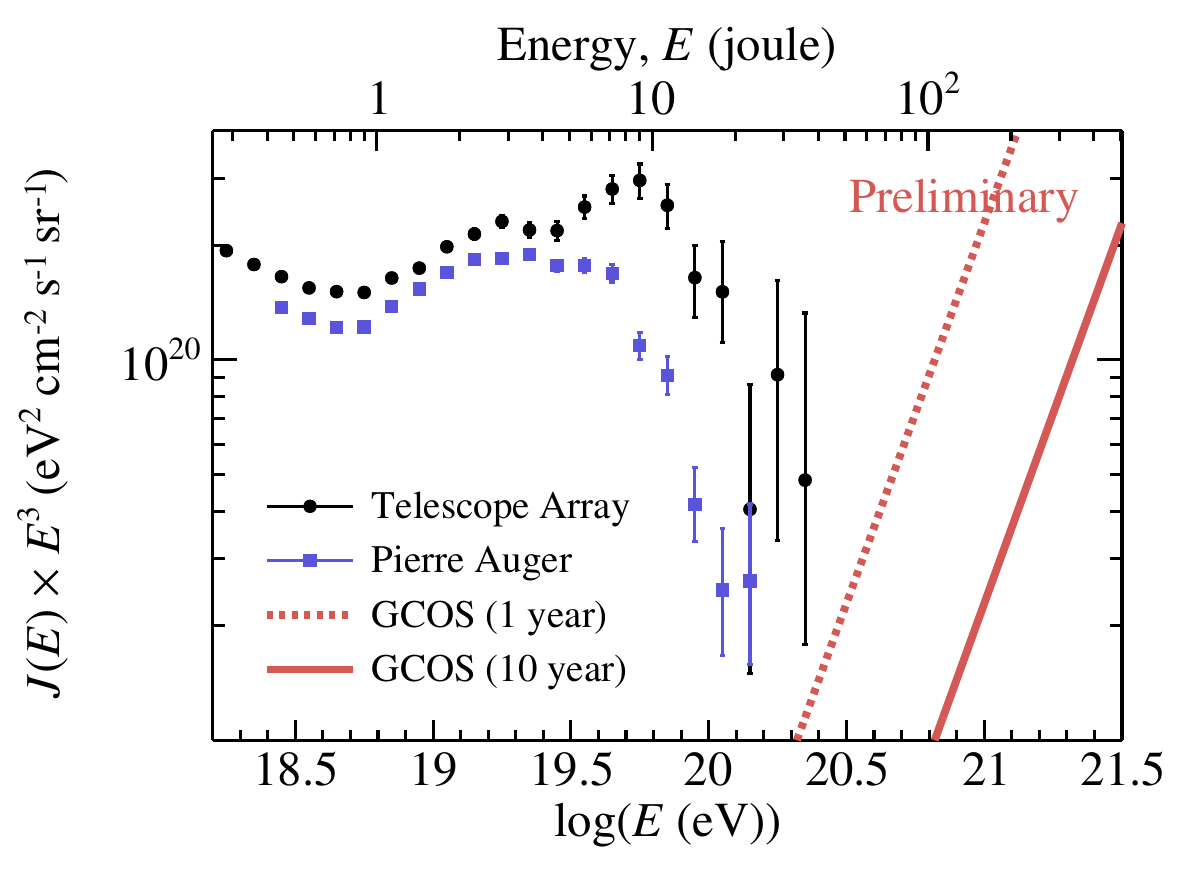}}
    \caption{(A) Expected total exposure with GCOS with a total effective area of 60,000\,km$^2$. Adopted from ~\cite{Ahlers:2025pqg}. (B) Expected 95\% confidence-level detection sensitivities with GCOS for one year and 10-years operations in comparison with energy spectra reported by Auger~\cite{PierreAuger:2020kuy,PierreAuger:2025hnw} and TA~\cite{Kim:2025yzr}.}
    \label{fig:gcos_performance}
\end{figure}

Figure~\ref{fig:gcos_performance}A shows an expected total exposure of GCOS with an effective area of 60,000 km$^2$ in comparison with current observatories~\cite{Coleman:2022abf}. 
The full-operation of GCOS will be expected to start after 2030 with an order of magnitude larger exposure than current observatories. 
Figure~\ref{fig:gcos_performance}B indicate the expected detection sensitivity of GCOS in case of one year and 10-year operations. 
This sensitivity is estimated from the effective area of 60,000\,km$^2$, a field-of-view of $\pi$\,steradian, and a detection threshold of three events corresponding to a 95\% confidence level.
GCOS is expected to accumulate statistics equivalent to those of existing observatories within one year,
and be a sensitive to search for the universe's most energetic particle with $\sim$1\,ZeV ($\equiv$ 10$^{21}$\,eV) in 10 years.
The scientific objectives of GCOS are described below. 
\begin{itemize}
    \item Establish charged-particle astronomy with UHECRs as a new messenger
    \vspace{-3mm}
    \item Clarify the mechanisms capable of accelerating particles to the highest energies in cosmic accelerators.
    \vspace{-3mm}    
    \item Constrain a detailed structure and strength of the Galactic magnetic field
    \vspace{-3mm}    
    \item Detect ultra-high-energy neutrinos and photons, and explore their synergies within a multi-messenger approach
    \vspace{-3mm}    
    \item Search for new physics beyond standard model, such as superheavy dark matter, magnetic monopole, exotic particle, and quantum gravity signatures such as Lorentz invariance violation
    \vspace{-3mm}    
    \item Understand hadronic interaction and air-shower physics beyond the energy region achieved by human-made particle accelerators
    \vspace{-3mm}    
    \item Study geophysics and earth science as interdisciplinary research in climate change
\end{itemize}

\section{Summary and future perspectives}
UHECRs are one of the important messenger to clarify extremely energetic phenomena in the universe.
The GCOS is the proposed next-generation observatories after 2030 for detecting UHECRs with the unprecedented exposure and mass identification capabilities. 
The total effective of GCOS is expected to be 60,000\,km$^2$ with accuracies of 1.0$^{\circ}$ for arrival direction, 10\% for energy, and $\Delta \ln(A)=1.0$ for mass number of primary particle.
GCOS will challenge the charged-particle astronomy to elucidate the origin and nature of UHECRs.
The optimizations of scientific objectives and detector designs are being continued and matured through the worldwide cooperation.

\section*{Acknowledgements}
This work was supported by JSPS KAKENHI Grant Number 25H00647, 21H04470, and JSPS Core-to-Core Program Grant Number JPJSCCA20250003.
This work was partially carried out by the joint research program of the Institute for Cosmic Ray Research (ICRR) at the University of Tokyo.
The author thanks to Pierre Auger and Telescope Array Collaborations, and the GCOS supporters for insightful and productive discussions.

\fontsize{10pt}{9pt}\selectfont
\bibliography{main}
\bibliographystyle{JHEP}

\end{document}